\documentclass[oneside,reqno]{article}
\usepackage{tabularx,longtable,booktabs,multirow,graphicx,lscape,amsmath}
\usepackage[numbers, square, sort&compress]{natbib}
\usepackage[nolist]{acronym}
\usepackage[normalem]{ulem}
\usepackage[usenames, dvipsnames]{color}
\usepackage[table, xcdraw]{xcolor}
\usepackage[margin=1.5in]{geometry}
\usepackage{hyperref}
\usepackage{soul}
\usepackage{rotating}
\usepackage{arydshln}
\usepackage{rotating}
\usepackage{setspace}
\usepackage{tcolorbox}

\definecolor{mycolor1}{RGB}{220,220,255}
\definecolor{mycolor2}{RGB}{0, 0, 225}

\newcommand{\insertitem}[1]{%
    \begin{center}
        \vspace{10pt} 
        [INSERT #1 HERE.]
        \vspace{10pt} 
    \end{center}
}
\newcommand{\hypothesis}[1]{
    \begin{center}
        \textit{#1}
    \end{center}
}

\begin{document}
\begin{acronym}[MPC]
    \acro{2SLS}{2-Stage Least Squares}
    \acro{AMI}{Acute Myocardial Infarction}
    \acro{CDHP}{Consumer-Driven Health Plans}
    \acro{CMS}{Centers for Medicare and Medicaid Services}
    \acro{VM}{Value-Based Payment Modifiers Program}
    \acro{COB}{Coordination of Benefits}
    \acro{DiD}{Difference-in-Difference}
    \acro{DRGs}{Diagnosis Related Groups}
    \acro{EP}{Eligible Professional}
    \acro{FDA}{Food and Drug Administration}
    \acro{FTC}{Federal Trade Commission}
    \acro{GNT}{Gowrisankaran, Nevo, and Town}
    \acro{GP}{General Practitioners}
    \acro{HHI}{Herfindahl-Hirschman Index}
    \acro{HMO}{Health Maintenance Organizations}
    \acro{HMOs}{Health Maintenance Organizations}
    \acro{HRR}{Hospital Referral Regions}
    \acro{HSA}{Health Service Areas}
    \acro{MarketScan data}{Truven Health MarketScan\textsuperscript{\textregistered} Research Databases}
    \acro{MSA}{Metropolitan Statistical Area}
    \acro{NHS}{National Health Service}
    \acro{OOP}{out-of-pocket}
    \acro{pdf}{probability density function}
    \acro{PPO}{Preferred-Provider Organizations}
    \acro{ProMedica}{ProMedica Health System, Inc.}
    \acro{SCM}{Synthetic Control Method}
    \acro{SCP}{Structure-Conduct-Performance}
    \acro{SD}{Standard Deviation}
    \acro{SLH}{St. Luke's Hospital}
    \acro{SOI}{Severity of Illness}
    \acro{Truven Health}{Truven Health Analytics\textsuperscript{\textregistered}, part of the IBM Watson Health\texttrademark business}
    \acro{UK}{United Kingdom}
    \acro{WTP}{willingness to pay}
    \acro{FE}{fixed effects}
    \acro{FD}{first differencing}
    \acro{RE}{random effects}
    \acro{MC}{mixed coefficients}
    \acro{IV}{instrumental variable}
    \acro{OLS}{Ordinary Least Squares}
    \acro{POLS}{Pooled Ordinary Least Squares}
    \acro{Wilcoxon test}{Wilcoxon rank distribution test}
    \acro{US}{United States}
    \acro{EU}{the European Union}
    \acro{HSR}{Hart-Scott-Rodino Act of 1976}
    \acro{DOJ}{Department of Justice}
    \acro{SSNIP}{small but significant non-transitory increase in price}
    \acro{IND}{Investigational New Drug}
    \acro{NDA}{New Drug Application}
    \acro{ANDA}{Abbreviated New Drug Application}
    \acro{ODA}{Orphan Drug Act}
    \acro{HAV}{Hepatitis A}
    \acro{HBV}{Hepatitis B}
    \acro{HCV}{Hepatitis C}
    \acro{HIV}{HIV/AIDS}
    \acro{BMS}{Bristol-Myers Squibb}
    \acro{GSK}{Glaxo Smith Kline}
    \acro{ViiV}{ViiV Healthcare}
    \acro{AWP}{Average Wholesale Price}
    \acro{Gilead}{Gilead Sciences}
    \acro{HDHP}{High-Deductible Health Plan}
    \acro{CDHP}{Consumer-Driven Health Plan}
    \acro{POS}{Point of Service}
    \acro{IP}{Intellectual Property}
    \acro{GAFAM}{Google, Amazon, Facebook, Apple, and Microsoft}
    \acro{IO}{Industrial Organization}
    \acro{IC}{Indifference Curve}
    \acro{ISP}{Internet Service Provider}
    \acro{Mbps}{megabit per second}
    \acro{AI/AN}{American-Indian/Alaskan-Native}
    \acro{MMA}{Modern Merger Analysis}
    \acro{CPI}{Consumer Price Index}
    \acro{OTC}{over-the-counter}
    \acro{JJ}{Johnson and Johnson}
    \acro{MS}{multiple sclerosis}
    \acro{CNS}{central nervous system}
    \acro{ADHD}{attention deficit hyper activity disorder}
    \acro{ATC}{Anatomical Therapeutic Chemical}
    \acro{NDC}{National Drug Code}
    \acro{HCP}{health care provider}
    \acro{HCF}{Healthcare Connect Fund}
    \acro{Telecom}{Telecommunications Program}
    \acro{RUCC}{Rural-Urban Continuum Code}
    \acro{FCC}{Federal Communications Commission}
    \acro{BO1}{Best of 1}
    \acro{BO3}{Best of 3}
    \acro{BO5}{Best of 5}
    \acro{CT}{counter-terrorist team}
    \acro{T}{terrorist team}
    \acro{CS:GO}{Counter Strike: Global Offensive}
    \acro{LAN}{Local Area Network}
    \acro{META}{Most Effective Tactic Available}
    \acro{AWP}{Arctic Warfare Police}
    \acro{MPLS}{Multiprotocol Label Switching}
    \acro{ISDN}{Integrated Services Digital Network}
    \acro{USAC}{Universal Service Administrative Company}
    \acro{QALY}{quality-adjusted life years}
    \acro{SPARCS}{Statewide Planning and Research Cooperative System}
    \acro{PCa}{prostate cancer}
    \acro{PSA}{prostate-specific antigen}
    \acro{CDC}{Centers for Disease Control and Prevention}
    \acro{NBS}{Nash Bargaining Solution}
    \acro{FOC}{first order condition}
    \acro{RHC}{Rural Health Care}
    \acro{USF}{Universal Service Fund}
    \acro{NIH}{National Institute of Health}
    \acro{FRC}{Federal Radio Commission}    
    \acro{GMM}{Generalized Method of Moments}
    \acro{CSR}{Corporate Social Responsibility}
    \acro{LBW}{Low Birth Weight}
    \acro{VLBW}{Very Low Birth Weight}
    \acro{ELBW}{Extremely Low Birth Weight}
    \acro{ER}{Emergency Room}
    \acro{SES}{socioeconomic status}
\end{acronym}

\title{Falling Birthrate and Rising C-section: Post-Pandemic Evidence from New York}

\author{Maysam Rabbani\thanks{Department of Economics, Feliciano School of Business, Montclair State University, NJ, USA.}
\and
Zahra Akbari\thanks{Department of Economics, Love School of Business, Elon University, Elon, NC, USA.}}

\maketitle

\begin{abstract}

\textbf{Background}: The literature documents the effects of the pandemic on birthrate, birthweight, and pregnancy complications. However, the evidence is mixed on the persistence of these effects post-pandemic.

\textbf{Objective}: This study examined whether birthrate has declined with the onset of the pandemic, whether the effects subsided post-pandemic, and whether there are discrepancies by race and ethnicity, or between vaginal delivery and C-section.

\textbf{Methods}: Using the 2012–2022 hospital inpatient discharge data from New York, we implemented fixed-effects regression models to estimate changes in birthrate and delivery method composition. The study was conducted at the zip code level and eliminated the effect of time-invariant characteristics such as geographic and demographic discrepancies.

\textbf{Results}: Birthrate had been declining pre-pandemic by 1.11\% annually. The onset of the pandemic in 2020 led to an additional 7.61\% decline in birthrate, which did not revert to the pre-pandemic trajectory in subsequent years, indicating a persistent decline. The post-pandemic decline was more pronounced in vaginal deliveries, with weak evidence of a drop in C-sections. In our sample, C-sections generate 61\% more revenue than vaginal deliveries, suggesting that healthcare providers may have increased C-section rates to offset revenue losses from declining birthrates.

\textbf{Conclusions}: The pandemic accelerated an ongoing decline in birthrate, which has persisted post-pandemic. The shift in delivery method composition raises concerns about potential financial incentives influencing clinical decisions. Further research is needed to confirm whether hospitals upsold C-sections in response to declining births.



\noindent \textbf{Highlights}:
\begin{itemize}
    \item Birthrate in New York State had been falling 1.11\% annually before the pandemic
    \item Birthrate dropped by an additional 7.61\% with the onset of the pandemic
    \item The effect of the pandemic persisted in the post-pandemic years
    \item The effect was stronger for vaginal delivery than C-section
\end{itemize}


\end{abstract}

\section{Introduction}
\label{sec:intro}

Years after the COVID-19 pandemic, researchers continue to uncover its lasting effects on the population. While it may be presumed that the shocks caused by the pandemic must have subsided post-pandemic, evidence of persistent effects continues to emerge \citep{rabbani_bw1}. Birthrate is among the areas that may face lasting consequences. Birthrate has been declining for over a century \citep{webb1907decline, perrott2005population, omram2001epidemiologic} across many geographies worldwide \citep{robey1993birth}, in the Middle East, North Africa \citep{pourreza2021contributing}, and China \citep{zhang2022analysis}. This decline is not a smooth trend. Two types of shocks are known to diminish birthrate: economic crises and pandemics.

A drop in birthrate has been documented after the 2008 market crash \citep{kearney2022puzzle}. Economic shocks can shake confidence in economic stability and growth, heighten concerns about unemployment, and weaken households' financial capacity to raise children \citep{sobotka2011economic}. Adverse economic shocks could reshape individual priorities, switching from long-term (building a family and having children) to short-term personal life aspirations \citep{kearney2022puzzle}.

Pandemics could strongly reduce birthrate. After each of the pandemics that occurred in 1890, 1918, 1920, and 1957, birthrate declined immediately, peaking 9 months into the pandemic and rebounding shortly after \citep{le2023remarkable}. There are several underlying mechanisms that explain this. Parents often seek reproductive healthcare before pregnancy to minimize birth complications. These services are harder to access during a pandemic \citep{verbiest2022listening, farley2007utilization}. Women have expressed concerns about getting pregnant during a pandemic \citep{aytha2022p61} due to elevated health and safety risks \citep{gartner2015calculating, tasneem2023fertility}. Pandemics could lower subjective well-being, increase stress levels, and exacerbate struggles in relationships, and these factors could reduce birthrate \citep{manning2022cognitive}. Pandemics also tend to hurt the economy, which is another pregnancy deterrent \citep{tasneem2023fertility}. 

Most studies conducted after the COVID-19 pandemic show a pattern that is in line with previous pandemics. Studies in Finland \citep{nisen2022fertility}, Wales and Australia \citep{gray2022having}, the \ac{US} \citep{kearney2023us}, the \ac{US} state of California \citep{bailey2023covid}, 24 European countries \citep{pomar2022impact}, and a group of 38 higher-income countries \citep{sobotka2024pandemic} show a sudden birthrate decline in 2020, followed by a swift recovery in 2021. However, a study in Spain found only a partial birthrate recovery after the pandemic \citep{fallesen2023partial}, and a study in Norway found that the pandemic led to an increase in birthrate \citep{lappegaard2024understanding}. A new and sharp decline in birthrate seems to have begun in 2022 and may be accelerating for no obvious reason \citep{winkler2024birth}. Birthrate dropped 10\% in Sweden and Germany in 2022, and this cannot be explained by factors such as unemployment rate, COVID-19 deaths, infection rates, or other socioeconomic characteristics. But it could be partially explained by vaccination programs \citep{bujard2024fertility, ma2023effects}. 

The objective of the current study was to enrich the literature by answering the following questions: (1) is there a persistent decline in birthrate before the COVID-19 pandemic? (2) is there a sudden birthrate decline concurrent with the pandemic? (3) and if the answer is yes, does birthrate after the pandemic revert to the pre-pandemic trajectory? (4) is there a post-pandemic discrepancy between vaginal delivery and C-section birthrate? And (5) are there birthrate discrepancies by race and ethnicity? Birthrate and pregnancy complications have far-reaching consequences for population well-being \citep{rabbani_bw1}. Therefore, it is an imperative health policy priority to understand the full extent of the pandemic's effect on birthrate, identify potential risk factors, and plan to respond accordingly.

By analyzing New York’s hospital inpatient discharges data during 2012-2022, we found a steady 1.11\% decline in birthrate before the pandemic, followed by a sudden 7.61\% decline with the onset of the pandemic. In the post-pandemic years, birthrate has not reverted to the pre-pandemic trajectory. This marked decline predominantly appeared in vaginal delivery cases, with no measurable impact on C-section. It is unclear if this shift towards C-section is the decision of mothers or it is an upselling attempt.

\section{Methods}
\label{sec:methods}

\subsection{Data}
\label{sec:data}
 
we employed the \ac{SPARCS} data. \ac{SPARCS} is New York state's hospital inpatient discharges data. Observations are at the level of the episode of care, i.e., there is one observation per each birth case. The data was de-identified. But socioeconomic characteristics were reported, including gender, race, ethnicity, and source of payment.

After data clean up (limiting to observations with full valid data for the outcome and explanatory variables) the data comprised a net sample of 2,392,800 observations, including 66.3\% vaginal and 33.7\% C-section births. To empirically test if there is a change in birthrate, we needed to aggregate the data at a reasonable level of geography and time. The observations were scattered across 51 3-digit ZIP codes in the state of New York. Therefore, we took each 3-digit ZIP code as one geographic unit of observation and counted the number of births in each unit per year. This generated 714 unique ZIP-year combinations.

\subsection{Hypotheses}
\label{sec:hypotheses}

Using the empirical specification that is described in section \ref{sec:statistical_model}, we tested the following four null hypotheses. 

\hypothesis{H1: birthrate was stable over time in the pre-pandemic years}

\textit{H1} forms a baseline for the analysis by testing if there was any systematic variation in birthweight in the years preceding the pandemic, once seasonal and demographic controls are applied. This forms what is often referred to as the counterfactual trend. To explain, if we establish no pre-pandemic trend, then we measure deviations during and after the pandemic as evidence of secular shocks caused by the pandemic. Alternatively, if we find a pre-pandemic trend, then we use that trend as the baseline, and we measure deviations from this trend as evidence of the effect of the pandemic.

\hypothesis{H2: birthrate declined with the onset of the pandemic}

As discussed, there is an ongoing body of literature on birthrate shocks caused by the pandemic, and assumption \textit{H2} allows us to place our study in this broader literature. Hence, \textit{H2} is a cornerstone of our analysis, where we test if the combination of health crisis and socioeconomic dynamics during the pandemic led to a measurable decline in births.

\hypothesis{H3: post pandemic, birthrate returned to the pre-pandemic trajectory}

Hypotheis \textit{H3} is where our study aims to set itself apart from existing literature, and test for the persistence of pandemic-era effects in the post-pandemic years. If we establish that the birth-rate shocks that began during the pandemic have reverted to the pre-pandemic trends, this would alleviate concerns about the long-term impacts of the pandemic. In contrast, if we establish that the shocks persisted post-pandemic, then the policy concerns would be substantially greater because long-lasting effects could have much costlier consequences for the health and longevity of the population.

\hypothesis{H4: the pandemic equally affected vaginal and C-section births}

One concern raised in the literature is that healthcare providers facing revenue declines may resort to upselling more profitable procedures to offset financial losses. C-sections are known to generate higher revenues than vaginal deliveries, while also being more predictable and requiring fewer hospital bed days per patient. This creates a potential incentive for providers to recommend C-sections as a means of compensating for reduced birth volumes, even in cases where vaginal delivery might otherwise be advised. \textit{H4} examines whether the COVID-19 pandemic, which potentially led to a decline in birth rates, also resulted in an increased proportion of C-section deliveries as a form of revenue recovery. By testing whether the pandemic affected vaginal and C-section births equally, \textit{H4} seeks evidence of upselling behavior in response to financial pressures.

The combination of the above hypotheses would shed light on the short-term and long-term effects of the pandemic on birthrate. It would also help uncover the potential presence of upselling efforts. 

\subsection{Statistical model}
\label{sec:statistical_model}

We employed two specifications to testing the hypotheses. In specification 1, we introduced a binary (dummy) control for the pandemic period, equal to $1$ in years 2020-2022 and $0$ otherwise. Another binary captured the  post-pandemic shifts: equal to $1$ in years 2021-2022 and $0$ otherwise. The combination of an annual time-trend and the above two binary indicators isolated the secular effect of the pandemic and post-pandemic \citep{rabbani_bw1}. In specification 2, we replaced year, pandemic, and post-pandemic measures by year \ac{FE}, where 2012 served as the omitted year, and 10 binaries captured annual changes during 2013-2022. We used specification 1 to conduct the regression analysis (Table \ref{tab:reg1}), whereas specification 2 enabled the event study (Figure \ref{fig:event}). The two specifications together painted a holistic picture of the phenomena. Specification 1 was as follows:

\begin{equation}
    \ln({Freq}_{zt}) = \beta_0 + \beta_1.Year_{zt} + \beta_2.Pandemic_{zt} + \beta_3.PostPandemic_{zt} + \epsilon_{zt}
    \label{eq:1}
\end{equation}

\noindent And specification 2 used the following structure \citep{rabbani_bb1, rabbani_bw1}:

\begin{equation}
    \ln({Freq}_{zt}) = \alpha_0 + \sum_{y=2013}^{2022}[\alpha_y.I^y_{zt}] + e_{zt}
    \label{eq:2}
\end{equation}

In the above models, $\ln({Freq}_{zt})$ is the natural logarithm of the number of birth cases in year $t$ and ZIP code $z$. The use of natural logarithm enhances the model's fit to normal distribution \citep{rabbani_bb1, rabbani_wtp1, kwoka2010price}, which is a fundamental assumption that underlies the model. $Year_{zt}$ is a continuous measure of time; $Pandemic_{zt}$ is equal to $1$ in 2020-2022, and $0$ otherwise; $PostPandemic_{zt}$ is equal to $1$ in 2021-2022, and $0$ otherwise; $I^y_{zt}$ is a binary indicator equal to $1$ in year $y\in [2013,2022]$ and $0$ otherwise; And $\epsilon_{zt}$ and $e_{zt}$ are the error terms with standard normal distributions. The standard errors were clustered at the ZIP code level to account for the heteroskedasticity that could arise due to variations across geographies.

In the regression models, we assigned a weight to each ZIP code proportionate to its average annual number of births. Not only did it enhance statistical power, but it also prevented inferences from being dominated by small, sparsely observed ZIP codes. It is important to note that the use of an \ac{FE} model automatically and fully controlled for all time-invariant factors including the ZIP code level socioeconomic composition (age, gender, race, ethnicity, income, etc.) to the extent that these factors remained stable over time.

\section{Results}
\label{sec:results}

Table \ref{tab:summary} reports the summary statistics. 75.2\% of the births took place in the pre-pandemic years, 8.3\% during the pandemic, and 16.5\% post pandemic. 33.7\% of the births opted for C-section. 50.6\% of the mothers were white, 15.2\% were black, and 34.2\% were either a different race or chose not to identify. 17.7\% of the sample were Hispanic, 75.8\% non-Hispanic, and 6.5\% chose not to identify. The patients were predominantly on Medicaid (47.7\%), private health insurance (26.6\%), or Blue Cross Blue Shield (20.5\%). The sample size is large and provides sufficient statistical power to reliably study the outcome measures.

\insertitem{TABLE 1}

\begin{figure}[ht]
    \centering
    \includegraphics[width=1\linewidth]{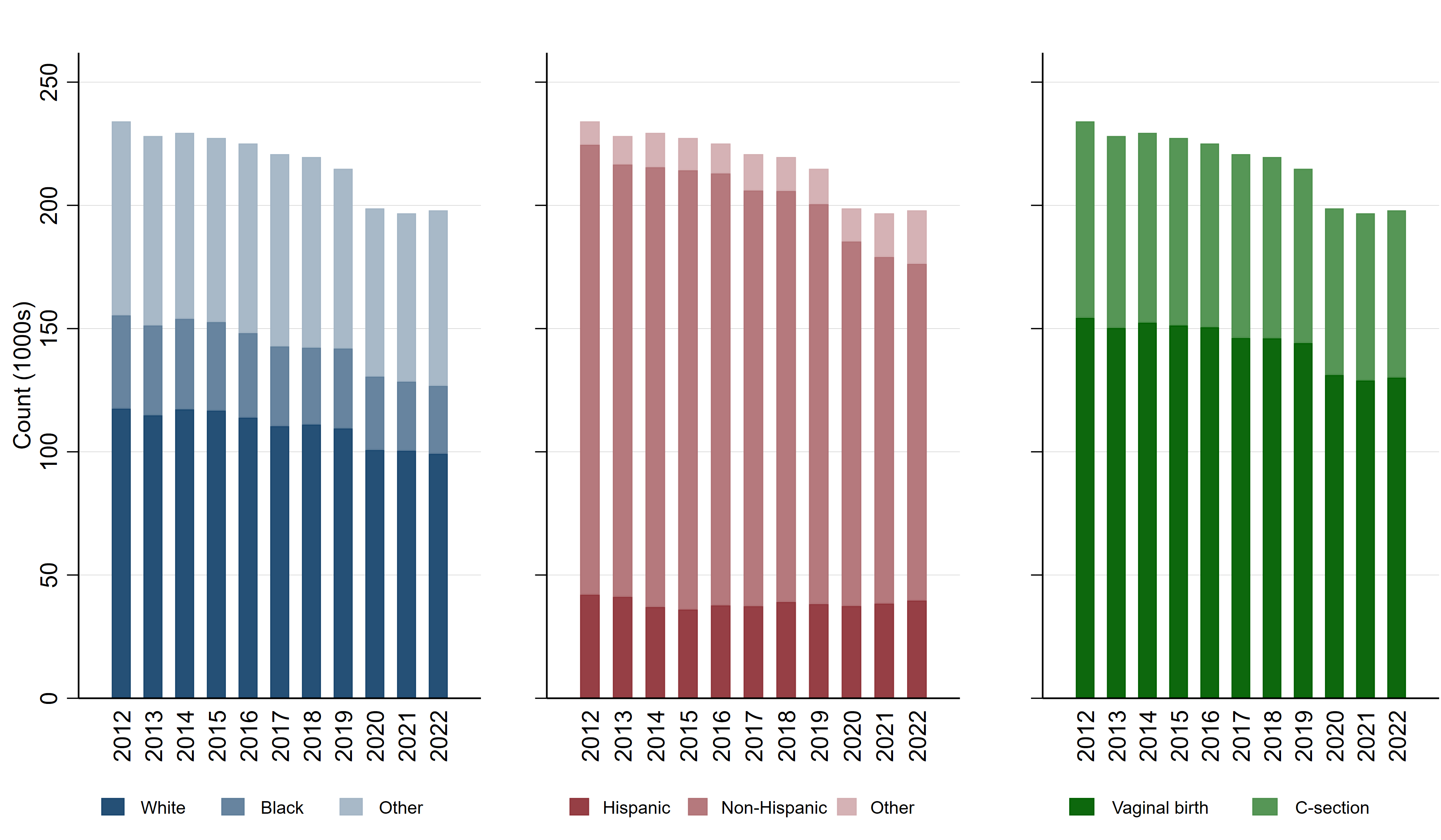}
    \caption{event study trends for the full sample, vaginal delivery, C-section, and by race and ethnicity.}
    \label{fig:raw_trend}
\end{figure}

Figure \ref{fig:raw_trend} illustrates the trends in the sample during 2012-2022. In each panel, the vertical axis measures the number of birth cases in thousands of newborns. The figures comprises three panels to separately show the sample composition by race (left), ethnicity (middle), and delivery modes (right).

Starting with race, three observations are noteworthy. First, the overall number of births is on a steady decline, starting with 234,080 cases in 2012, and ending at 197,970 cases in 2022. This is consistent with what was anticipated based on existing literature. Second, there is a sudden drop in birth count in 2020 and it seems to have not recovered in 2021 nor in 2022. Third, there is a potential racial difference in terms of the amount of decline in births with the onset of the pandemic. 

The ethnicity trends suggest that the sudden decline with the onset of the pandemic is not equally affecting all groups. Specifically, there is no visible decline among Hispanic people, a visible decline in non-Hispanic, and an increase in births among people who identified as ``other'' ethnicities.

Regarding delivery modes, both groups experienced a decline: vaginal delivery fell from 154,378 cases in 2012 to 130,149 in 2022, and C-section fell from 79,702 in 2012 to 67,821 cases in 2022. It implies 15.7\% fewer vaginal and 14.9\% fewer C-section births. We emphasize that no conclusion could be drawn by inspecting the trends, and we leave all inferences to be made based on the regression estimates in section \ref{sec:results}.

\insertitem{FIGURE 1}

Table \ref{tab:reg1} reports the baseline regression estimates using specification 1. All models employed \ac{FE} regression. Standard errors, clustered at the ZIP code level, are reported in parenthesis. Each column reports the coefficients for three main variables: $Year$ measures the annual change in birthrate before the pandemic; $Pandemic$ measures the additional change in birthrate that occurred with the onset of the pandemic; and $PostPandemic$ measures the additional post-pandemic change. If the coefficient of $PostPandemic$ is small and insignificant, it means that the effect of the pandemic has persisted post pandemic. Conversely, if the coefficient of $PostPandemic$ is the negative of the coefficient of $Pandemic$, it would indicate a full recovery to the pre-pandemic levels.

\insertitem{TABLE 2}

Starting with column 1, birthrate was falling 1.11\% annually pre pandemic, and it dropped by an additional 7.61\% when the pandemic hit. The coefficient for post pandemic is insignificant and close to zero, indicating that the decline in the birthrate that began with the pandemic has persisted post pandemic.

Columns 2 and 3 provide a comparison of the differences between vaginal delivery and C-section. Vaginal delivery and C-section were both falling pre-pandemic at annual rates of 0.92\% and 1.49\%, respectively. However, with the onset of the pandemic, vaginal delivery rate dropped twice the C-section rate (9.12\% versus 4.71\%). Post pandemic, vaginal delivery did not recover whereas C-section rate grew 2.02\%. This indicates that vaginal delivery was affected more than C-section during the pandemic, and vaginal delivery fell further behind C-section post pandemic. 

A comparison by race (columns 4-6) revealed that, pre pandemic, birthrate was falling much faster for black (2.92\%) than white (1.02\%) race. When the pandemic hit, birthrate declined more (7.65\%) for white than black (2.90\%) patients. Post pandemic, there was no birthrate change for white, but black patients experienced an additional 3.15\% decline. So, in the sum of pandemic and post pandemic, white and black patients had a comparable total decline in birthrate (7.76\% versus 6.05\%). Other/unknown races showed no pre-pandemic decline, had a 10.3\% decline during the pandemic, and did not recover post pandemic. One must read the results by race with caution because any white or black patient who chose not to identify was placed under ``other/unknown'' race. In other words, there was a potentially strong selection bias in the results by race that would hinder inference.

Columns 7-9 report the results for Hispanic, non-Hispanic, and other/unknown ethnicities. The results are intriguing: non-Hispanic patients experienced a 1.44\% annual decline in birthrate before the pandemic, a 9.39\% decline when the pandemic hit, and an additional 5.42\% decline post pandemic. Meanwhile, for Hispanic patients, birthrate was not falling pre pandemic nor during the pandemic, and rising by 5.30\% post pandemic. While this may indicate a discrepancy by ethnicity, the results could be contaminated by patients' decision to identify their ethnicity. 

\begin{figure}[ht]
    \centering
    \includegraphics[width=1\linewidth]{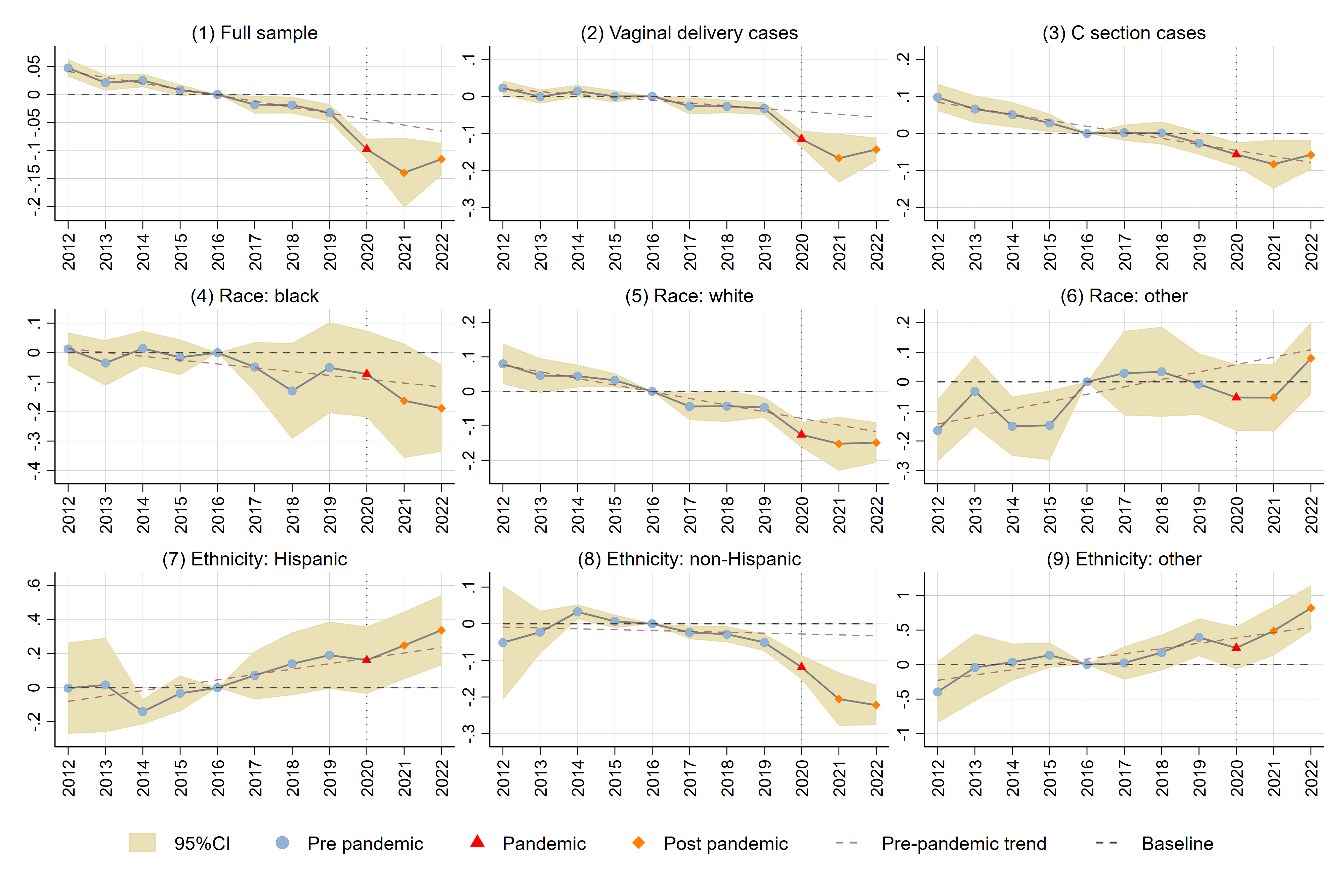}
    \caption{event study trends for the full sample, vaginal delivery, C-section, and by race and ethnicity.}
    \label{fig:event}
\end{figure}

Figure \ref{fig:event} illustrates the event study results using specification 2. The pandemic period is marked with a red triangle, post pandemic shown with yellow diamonds, and pre pandemic is marked with blue circles. A 95\% confidence interval envelops the trend lines to facilitate interpretation of statistical significance. 

\insertitem{FIGURE 2}

The vertical dotted line in year 2020 marks the pandemic. The horizontal dashed black line is the baseline for reference. If birthrate is stable over time, the trends must closely follow the black line. In contrast to the black line, the dashed brown line shows the pre-pandemic trend of birthrate. The brown line facilitates interpretation in two ways. First, the deviation of the brown line from the black line visualizes the pace of decline. Second, the post-pandemic deviation of the trend below the brown line demonstrates the extent of the sudden drop in birthrate during and after the pandemic. For ease of comparison, the 9 panels in Figure \ref{fig:event} appear in the same order as models 1-9 in Table \ref{tab:reg1}.

Panel 1 shows a steady and sharp decline in birthrate before the pandemic. It also corroborates that as the pandemic hit, the trend dropped below the pre-pandemic trajectory (brown line). Although there seems to be a partial post-pandemic recovery, the trend remained below the brown line, indicating a persistent decline.

Turning to panels 2 and 3, vaginal delivery follows a similar evolution over time as the baseline model, but there is no discernible effect of the pandemic on C-section. This corroborates that the effect of the pandemic is either exclusive to vaginal delivery or it affected vaginal delivery more severely than C-section. 

As shown in panels 4-6, the trends by race are less compelling both in terms of statistical significance and economic magnitude. Therefore, the event study does not support a racial discrepancy. Panels 7-9 partially support the idea that the effect was stronger on non-Hispanic patients. But for the caveats explained above, the evidence for an ethnic discrepancy remains weak and inconclusive.

\section{Discussions, limitations, and conclusions}
\label{sec:discussions}

The results could be summarized into a few key findings. First, the results support the existing literature \citep{webb1907decline, perrott2005population, omram2001epidemiologic} that birthrate is on a steady and gradual decline path. This observation is of high policy importance, and future work is warranted to understand the root causes, potential long-term consequences, and adopting suitable policies to revert it.

Second, we confirmed existing literature \citep{le2023remarkable, nisen2022fertility, gray2022having, kearney2023us, bailey2023covid, pomar2022impact, sobotka2024pandemic} that birthrate has indeed declined as the pandemic hit. However, we cast doubt on the broad consensus that birthrate has reverted to pre-pandemic levels after the pandemic. We found that the birthrate decline that began with the pandemic persisted after the pandemic. Our results lend support to the few studies who found that birthrate has not recovered \citep{fallesen2023partial} or even exacerbated \citep{winkler2024birth, bujard2024fertility} post pandemic.

Third, the results suggested that vaginal delivery was affected more than C-section, and that there was a partial recovery of C-section rate post pandemic that did not appear in vaginal births. While birthrate was falling during and after the pandemic, more patients underwent C-section. We conjecture that one of the following factors could be at play. First, C-section often entails a shorter hospital stay in a more controlled environment. So, pregnant women may have opted for C-section to reduce their exposure and risk during the peak of the pandemic. This could explain the rise of C-section utilization during the pandemic. But it may not explain a persistently higher C-section rate post pandemic. 

The second explanation is C-section upselling. In the sample, C-section generated 61\% more revenue per patient than vaginal delivery. Arguably, a decline in birthrate would cut into provider revenues. To boost revenues, providers may have attempted to upsell C-section to patients who would otherwise be advised to have a vaginal delivery. C-section upselling is a well-understood phenomenon \citep{gruber1999physician, rabbani2021non, chen2014failure}. A \$1,000 increase in the reimbursement for C-section delivery is estimated to lead to a 1\% increase in C-section use \citep{grant2009physician}. It is challenging to measure upselling because a change in mothers' preferences could lead to a change in C-section rate as well \citep{ma2010declining}. Therefore, it would be a worthy endeavor for future research to uncover the possibility of upselling in this context. 

The study was limited in the following ways. First and foremost, this paper was a partial attempt at causally linking the pandemic to birthrate. To explain, the use of an \ac{FE} specification convincingly controlled all time-invariant latent factors, and in this sense, it lent support to the causality of the findings. However, if there were time-variant latent changes (regulatory and reimbursement reforms \citep{jeungimproving, webster2022state}, maternal socioeconomic and behavioral characteristics \citep{perkowski2024co21, bolbocean2022ee596, zhu2016prevalence, smolen2015development, hobbs2023change, bolster2025association, siddika2023impact, yu2020birth, moran2020predictors}, or migration patterns) that coincided with the pandemic, it could confound and bias the study. This is a limitation that most existing studies share. Second, the data did not report the exact delivery date. Therefore the assumption, that year 2020 corresponded to the pandemic and years 2021-2022 corresponded to post pandemic, remains an imperfect match to the true start and end date of the pandemic. Third, the study was limited to the state of New York. So, the generalizability of the findings must be tested in future research. Nevertheless, New York is a high-population area with high economic and political importance, and understanding the dynamics that shape its healthcare would be an impactful policy endeavor.

\section{Implications for Behavioral Health and Future Research}

The persistent post-pandemic drop in birthrates, especially in vaginal deliveries, has significant behavioral health implications. Increased stress, uncertainty, and changes in healthcare-seeking behaviors during and after the pandemic may have deeply influenced reproductive choices and family planning decisions. The disproportionate increase in C-sections—even as overall births decline—suggests that medical decisions may be driven more by hospital financial pressures than patient needs, potentially eroding maternal agency and confidence in healthcare systems. Future studies should examine the psychological and social factors behind changing delivery preferences, including fear-driven choices, shifts in doctor-patient relationships, and the mental health impact of pregnancies during the pandemic. Additionally, it is essential to assess how financial motives influence unnecessary C-sections and whether these practices worsen inequities in maternal health outcomes.





\pagebreak



\pagebreak

\begin{table}[h]
\centering
\caption{summary statistics.}
\label{tab:summary}
\resizebox{.7\columnwidth}{!}{%
\begin{tabular}{lcccccc}
\hline
Variable & Obs & Mean & SD & Min & Median & Max \\ \hline
C-section & 2,392,800 & 0.337 & 0.473 & 0.000 & 0.000 & 1.000 \\
Pandemic (2020-2022) & 2,392,800 & 0.248 & 0.432 & 0.000 & 0.000 & 1.000 \\
Post pandemic   (2021-2022) & 2,392,800 & 0.165 & 0.371 & 0.000 & 0.000 & 1.000 \\
Year & 2,392,800 & 2016.82 & 3.140 & 2012 & 2017 & 2022 \\
\textbf{Race} &  &  &  &  &  &  \\
\textit{White} & 2,392,800 & 0.506 & 0.500 & 0.000 & 1.000 & 1.000 \\
\textit{Black} & 2,392,800 & 0.152 & 0.359 & 0.000 & 0.000 & 1.000 \\
\textit{Other} & 2,392,800 & 0.342 & 0.474 & 0.000 & 0.000 & 1.000 \\
\textbf{Ethnicity} &  &  &  &  &  &  \\
\textit{Hispanic} & 2,392,800 & 0.177 & 0.382 & 0.000 & 0.000 & 1.000 \\
\textit{Non-Hispanic} & 2,392,800 & 0.758 & 0.428 & 0.000 & 1.000 & 1.000 \\
\textit{Other} & 2,392,800 & 0.065 & 0.247 & 0.000 & 0.000 & 1.000 \\
\textbf{Insurance   type} &  &  &  &  &  &  \\
\textit{Medicaid} & 2,392,800 & 0.477 & 0.499 & 0.000 & 0.000 & 1.000 \\
\textit{Medicare} & 2,392,800 & 0.006 & 0.077 & 0.000 & 0.000 & 1.000 \\
\textit{Private health   insurance} & 2,392,800 & 0.266 & 0.442 & 0.000 & 0.000 & 1.000 \\
\textit{Blue Cross   Blue Shield} & 2,392,800 & 0.205 & 0.403 & 0.000 & 0.000 & 1.000 \\
\textit{Managed care} & 2,392,800 & 0.023 & 0.150 & 0.000 & 0.000 & 1.000 \\
\textit{Self-pay} & 2,392,800 & 0.011 & 0.103 & 0.000 & 0.000 & 1.000 \\
\textit{Federal, State,   Local, or VA} & 2,392,800 & 0.011 & 0.103 & 0.000 & 0.000 & 1.000 \\
\textit{Other} & 2,392,800 & 0.002 & 0.046 & 0.000 & 0.000 & 1.000 \\ \hline
\end{tabular}%
}
\end{table}

\begin{table}[h]
\centering
\caption{Regression estimates.}
\label{tab:reg1}
\resizebox{\columnwidth}{!}{%
\begin{tabular}{llllllllll}
\hline
 & \multicolumn{3}{c}{\cellcolor[HTML]{CBCEFB}Delivery method} & \multicolumn{3}{c}{Race} & \multicolumn{3}{c}{\cellcolor[HTML]{CBCEFB}Ethnicity} \\
\multirow{-2}{*}{Demographic} & \multicolumn{1}{c}{All} & \multicolumn{1}{c}{Vaginal} & \multicolumn{1}{c}{\begin{tabular}[c]{@{}c@{}}C\\ Section\end{tabular}} & \multicolumn{1}{c}{White} & \multicolumn{1}{c}{Black} & \multicolumn{1}{c}{\begin{tabular}[c]{@{}c@{}}Other/\\ unknown\end{tabular}} & \multicolumn{1}{c}{Hispanic} & \multicolumn{1}{c}{\begin{tabular}[c]{@{}c@{}}Non\\ Hispanic\end{tabular}} & \multicolumn{1}{c}{\begin{tabular}[c]{@{}c@{}}Other/\\ unknown\end{tabular}} \\
 & \multicolumn{1}{c}{(1)} & \multicolumn{1}{c}{(2)} & \multicolumn{1}{c}{(3)} & \multicolumn{1}{c}{(4)} & \multicolumn{1}{c}{(5)} & \multicolumn{1}{c}{(6)} & \multicolumn{1}{c}{(7)} & \multicolumn{1}{c}{(8)} & \multicolumn{1}{c}{(9)} \\ \hline
Year & -0.0111*** & -0.00919*** & -0.0149*** & -0.0102*** & -0.0292*** & -0.00213 & -0.00627 & -0.0144*** & 0.150** \\
 & (0.00234) & (0.00269) & (0.00254) & (0.00373) & (0.00669) & (0.00900) & (0.0113) & (0.00386) & (0.0716) \\
Pandemic & -0.0761*** & -0.0912*** & -0.0471*** & -0.0765*** & -0.0290 & -0.103*** & 0.00366 & -0.0939*** & -0.493** \\
 & (0.00745) & (0.00842) & (0.0110) & (0.0137) & (0.0287) & (0.0120) & (0.0380) & (0.0146) & (0.192) \\
PostPandemic & 0.00341 & -0.00543 & 0.0202 & -0.00112 & -0.0315* & 0.0155 & 0.0530** & -0.0542*** & 0.178 \\
 & (0.0100) & (0.00960) & (0.0142) & (0.0179) & (0.0178) & (0.00930) & (0.0209) & (0.0183) & (0.125) \\
Intercept & 31.43*** & 27.14*** & 37.88*** & 28.70*** & 66.56*** & 12.63 & 20.43 & 37.72*** & -296.7** \\
 & (4.715) & (5.427) & (5.124) & (7.513) & (13.48) & (18.15) & (22.84) & (7.783) & (144.3) \\
R2 & \multicolumn{1}{c}{0.504} & 0.490 & 0.420 & 0.335 & 0.368 & 0.092 & 0.006 & 0.431 & 0.182 \\
Obs & 2,368,850 & 1,569,117 & 799,733 & 1,202,729 & 357,632 & 808,555 & 418,175 & 1,800,117 & 150,582 \\  \hline
\end{tabular}%
} \\
{\raggedright
\fontsize{6}{6.5}\selectfont 
The dependent variable is the natural logarithm of the number of births at the level of ZIP code and year. Standard errors clustered at the county level are reported in parenthesis. *** $p<0.01$, ** $p<0.05$, * $p<0.1$. \par}
\end{table}

\pagebreak

\end{document}